\documentclass[twocolumn,aps,floatfix,superscriptaddress]{revtex4}
\usepackage{amsmath,amssymb,eucal,graphicx,float,epstopdf}
\begin{document}
\title{Extinction and Survival in Two-Species Annihilation}
\author{J.~G.~Amar} 
\affiliation{Department of Physics and Astronomy, 
University of Toledo, Toledo, Ohio 43606, USA}
\author{E.~Ben-Naim}
\affiliation{Theoretical Division and Center for Nonlinear Studies,
Los Alamos National Laboratory, Los Alamos, New Mexico 87545, USA}
\author{S.~M.~Davis} 
\affiliation{Department of Physics and Astronomy, 
University of Toledo, Toledo, Ohio 43606, USA}
\author{P.~L.~Krapivsky}
\affiliation{Department of Physics, Boston University, Boston,
  Massachusetts 02215, USA}
\begin{abstract}
We study diffusion-controlled two-species annihilation with a finite
number of particles.  In this stochastic process, particles move
diffusively, and when two particles of opposite type come into
contact, the two annihilate.  We focus on the behavior in three
spatial dimensions and for initial conditions where particles are
confined to a compact domain.  Generally, one species outnumbers the
other, and we find that the difference between the number of majority
and minority species, which is a conserved quantity, controls the
behavior.  When the number difference exceeds a critical value, the
minority becomes extinct and a finite number of majority particles
survive, while below this critical difference, a finite number of
particles of both species survive.  The critical difference $\Delta_c$
grows algebraically with the total initial number of particles $N$,
and when $N\gg 1$, the critical difference scales as $\Delta_c\sim
N^{1/3}$.  Furthermore, when the initial {\em concentrations} of the
two species are equal, the average number of surviving majority and
minority particles, $M_+$ and $M_-$, exhibit two distinct scaling
behaviors, $M_+\sim N^{1/2}$ and $M_-\sim N^{1/6}$.  In contrast, when
the initial {\em populations} are equal, these two quantities are
comparable $M_+\sim M_-\sim N^{1/3}$.
\end{abstract}

\maketitle

\section{Introduction}

Theoretical studies of non-equilibrium dynamics are primarily
concerned with the behavior of infinitely extended systems
\cite{krb,ut}. Indeed, the statistical physics of time-dependent
phenomena such as ordering \cite{krb,ut,ajb}, avalanches
\cite{dd,sdm,pld} and reaction processes \cite{otb,bh,thv} typically
focus on scaling laws for unbounded systems composed of infinitely
many interacting particles (or spins).  In most cases, theoretical
techniques which successfully describe infinite systems, cannot be
specialized to finite ones \cite{kpwh,fds}.  Yet, experimental
\cite{ak,dbm,se} and computational \cite{nb} studies necessarily
involve finite systems.

Reaction-diffusion processes (see \cite{krb,otb,bh,thv} for a review)
constitute an important class of non-equilibrium dynamics
\cite{cg}. Recent studies show that these processes exhibit phenomena
that are unique to finite systems \cite{bk,bk1}.  In particular, for
reaction-controlled single-species annihilation, it was recently found
that a finite number of particles may survive
indefinitely. Specifically, starting with a finite number of particles
which are confined to a bounded domain, a small subset of particles
may ``escape'' far outside the initially occupied region and thereby
avert annihilation. Here, we study two-species annihilation and find
another phenomena, a transition from survival to extinction, together
with scaling laws that are unique to finite systems.
                 
We investigate a random process where two distinct species diffuse in
unbounded space and additionally, the two species annihilate each
other.  While we also present results for general spatial dimensions,
we focus primarily on the most interesting case of three dimensions.
When a finite number of particles is initially localized within a
compact domain, there are two greatly different outcomes. In the first
scenario, a finite number of particles of each species survives the
annihilation process. In the second scenario, one species vanishes and
one species partially survives.  The initial population difference
controls this transition from survival to extinction.

Generally, one species outnumbers the other.  Moreover, the difference
between the number of majority particles and the number of minority
particles is conserved throughout the annihilation process. This
population difference controls the outcome of the reaction process and
moreover, there is a critical difference $\Delta_c$.  When the number
difference exceeds the critical difference, all minority particles
eventually vanish, but in the complementary regime, some minority
particles do survive.

A number of finite-size scaling laws characterize these behaviors.
First, the critical difference grows algebraically with the total
number of particles $N$,
\begin{equation}
\label{critical}
\Delta_c \sim N^{1/3}\,.
\end{equation}
We investigate the total number of surviving particles from each
species, and we consider two initial conditions: (i) equal populations
of the two species, and (ii) equal concentrations of the two species.
For equal populations, the behavior is very similar to that found for
single-species annihilation \cite{bk}. In this case, the average
number of surviving particles $M$ grows algebraically with system
size,
\begin{equation}
\label{ep}
M\sim N^{1/3}\,.
\end{equation}
For equal concentrations, the average number of surviving majority
particles, $M_+$, is much larger than the average number of surviving
minority particles, $M_-$. Interestingly, these two averages exhibit
different scaling laws,
\begin{equation}
\label{ec}
M_+\sim N^{1/2},\qquad M_- \sim N^{1/6}\,. 
\end{equation}
Neither one of these two scaling behaviors coincides with
Eq.~\eqref{ep}. Our theoretical analysis combines the rate equation
approach with scaling estimates for the {\em finite} duration of the
reaction process. Results of extensive numerical simulations confirm
the theoretical predictions.

\section{Two-Species Annihilation}

In the two-species annihilation process, particles are initially
distributed randomly in space with a uniform concentration. There are
two types of particles, denoted by $A$ and $B$. Each particle moves
diffusively, the diffusion coefficient $D$ is assumed to be the same
for both species.  Particles of the same type do not interact but two
particles of the opposite type annihilate upon contact, as represented
by the reaction scheme
\begin{equation}
\label{process}
A+B\to \emptyset\,.
\end{equation}
This stochastic process can be realized in continuous or discrete
space.  Our numerical simulations implement the discrete version where
particles occupy sites of a regular lattice.  Each particle performs a
random walk as it moves from one lattice site to a randomly-chosen
neighboring site. Annihilation occurs whenever a particle lands on a
site that is occupied by a particle of the opposite type. Two-species
annihilation has been used to model bimolecular chemical reactions
\cite{otb,se}, particle-antipaticle annihilation in the early universe
\cite{tw}, and particle-hole recombination in irradiated
semiconductors \cite{ak}. 

Two-species annihilation has been studied extensively for unbounded
systems populated by infinitely many particles, typically starting
with equal concentrations of the two species.  The spatial dimension
$d$ controls the behavior and there are two regimes. In sufficiently
low spatial dimensions, $d<4$, $A$-rich domains and $B$-rich domains
develop and since spatial correlations are
significant, the particle concentration $c$ decays slowly with time
$t$, namely $c\sim t^{-d/4}$ \cite{ybz,bur,zo,boo,dba,bl,gr,br,lr,ik,ea,hvb,rd}. In
sufficiently large dimensions, $d>4$, spatial correlations do not play
a significant role, and the concentration decays more rapidly, $c\sim
t^{-1}$. In the latter case, the decay exponent is universal and
further, the prefactor does not depend on the initial concentration.

Here we study the annihilation process \eqref{process} when the number
of particles is {\em finite}. Initially, $N$ particles are randomly
distributed inside a bounded domain \cite{bms}, which is embedded in
infinite space. Without loss of generality, we set the initial
concentration to unity such that the volume of the domain $V$ equals
the number of particles, $V=N$. In the simulations, we used spherical
domains for the initial configuration.  This set-up mimics physical
processes such as the recombination of vacancies and interstitials
produced in crystals by neutron, ion or electron radiation \cite{dbm}.

A recent study \cite{bk} of a diffusion-controlled annihilation
process involving a single type of particles has shown that starting
with a finite number of particles, on average, a finite number of
particles avoid annihilation as these surviving particles ``escape''
far outside the initially confining domain. Our goal is to study this
escape phenomena when there are two species.

While in the case of equal concentrations the populations of both
species are equal on average, for a given realization, one population
is larger than the other.  Let $N_+$ be the initial majority
population and $N_-$ be the initial minority population. The total
initial population is $N=N_++N_-$, and we consider the case where the
minority constitutes a finite fraction of the population $N_-\propto
N$. The population difference $\Delta$ is defined as
\begin{equation}
\label{Delta}
\Delta=N_+-N_-\,.
\end{equation}
Each annihilation event decreases the number of majority and minority
particles by one and therefore, the population difference is a
conserved quantity.

We denote the average number of surviving majority (minority)
particles by $M_+$ ($M_-$).  Conservation of the population difference
implies $M_+-M_-=\Delta$. As long as the two initial populations are
not equal, $\Delta>0$, some majority particles do survive, $M_+\geq
\Delta$.

However, there is no guarantee that minority particles survive, and
indeed, in sufficiently small dimensions, all minority particles are
annihilated. In dimension $d\leq 2$, a random walk is recurrent as it
is guaranteed to return to its starting position \cite{mp}.  Since
each particle performs a random walk, the distance between any two
particles itself performs a random walk. Hence, even if a minority
particle survives to a very late time, it is bound to eventually
encounter a majority particle. This argument shows that in spatial
dimensions $d\leq 2$, the minority species becomes extinct while the
number of surviving majority particles is deterministic, $M_-=0$ and
$M_+=\Delta$.  We now consider the behavior in three dimensions.

\section{Rate Equations} 

Our approach generalizes the methods previously used to analyze the
one-species annihilation process \cite{bk}. We assume that particles
are confined to a domain with volume $V$ and that they are uniformly
distributed inside this region. Ignoring spatial correlations, the
average numbers of majority and minority particles, $n_+(t)$ and
$n_-(t)$, at time $t$, obey the rate equations
\begin{equation}
\label{rate-eq}
\frac{dn_+}{dt} = - \frac{n_+ n_-}{V}\,, \qquad
\frac{dn_-}{dt} = - \frac{n_+ n_-}{V}\,.
\end{equation}
Without loss of generality, we set the reaction rate to
unity. Equation \eqref{rate-eq} can be derived from the rate equations
for the concentrations $c_+$ and $c_-$ inside the occupied domain with
volume $V$: we simply substitute $c_+=n_+/V$ into the canonical rate
equation $dc_+/dt=-c_+c_-$ \cite{krb,mvs}. By subtracting one
equation in \eqref{rate-eq} from the other, we confirm that the
population difference, $n_+-n_-$, is conserved.

As shown in \cite{bk}, there are two regimes of behavior. At early
times, particles remain inside the initially confining region with
volume $V=N$. At late times, particles manage to diffuse outside the
initial domain but are confined to an expanding region whose linear
dimension grows diffusively with time. Hence,
\begin{equation}
\label{V}
V(t) \sim 
\begin{cases}
N           & \quad t \ll T;\\
t^{3/2}      & \quad t \gg T.
\end{cases}
\end{equation}
The crossover time $T$ can be estimated by matching the two behaviors, 
\begin{equation}
\label{T}
T\sim N^{2/3}\,.
\end{equation}
The quantity $T$ is simply the diffusion time, $T\sim L^2$, that
characterizes the time it takes a particle to exit the initially
occupied domain with linear size $L\sim N^{1/3}$.

\begin{figure}[t]
\includegraphics[width=0.45\textwidth]{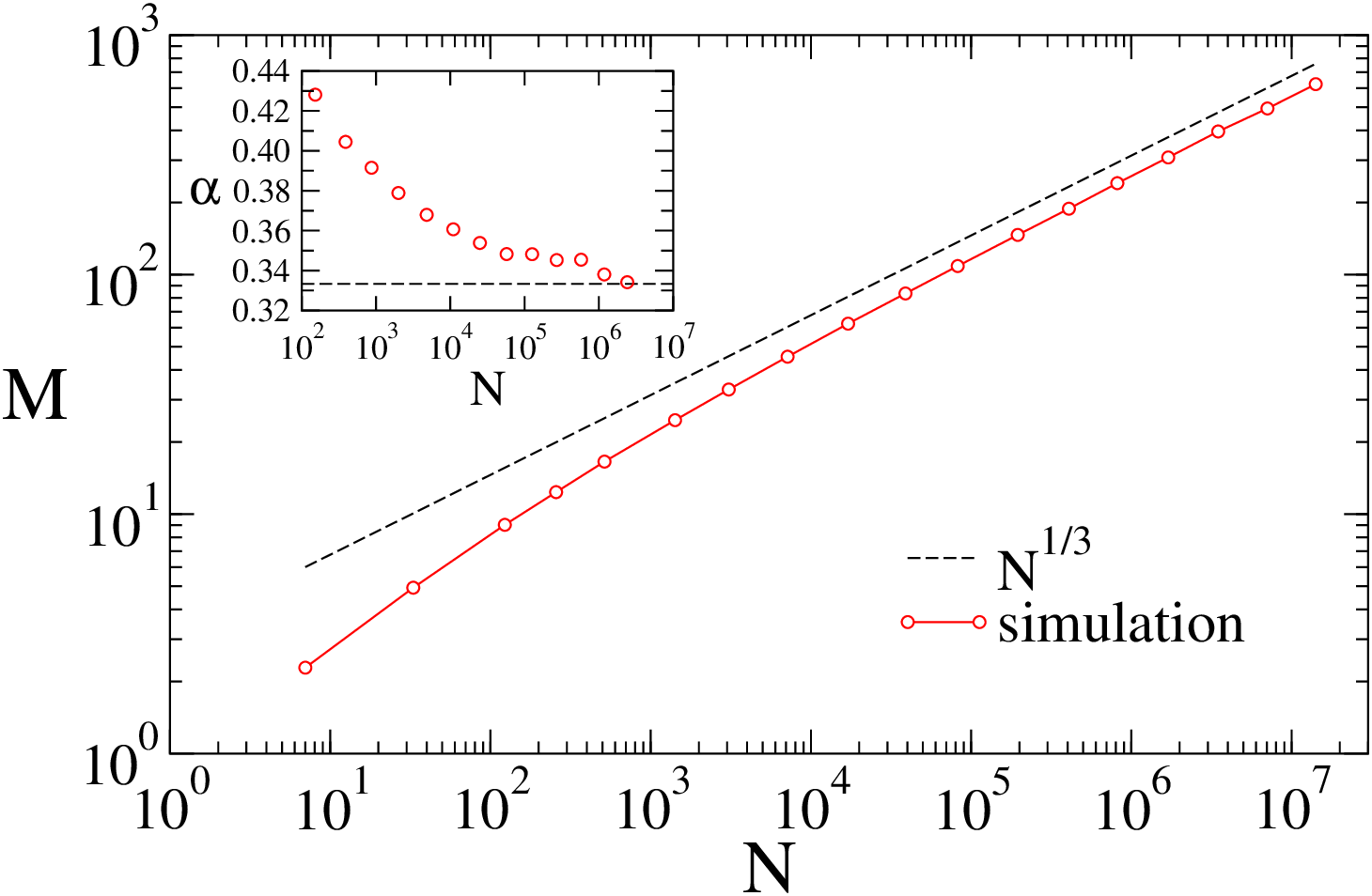}
\caption{The average number of surviving particles $M$ versus system
  size $N$ for equal populations. The inset shows the quantity
  $\alpha\equiv d\ln M/d\ln N$ versus $N$. Fitting the data to the
  powerlaw $M\sim N^\alpha$ in the range for $N>10^4$ yields the
  exponent $\alpha=0.34$.}
\label{fig-mnep}
\end{figure}

For single-species annihilation, it was found that the bulk of the
reaction events occur during the early phase. Furthermore, while rare
additional annihilation events may occur in the late phase, the number
of such reactions does not alter the scaling laws for the ultimate
number of surviving particles. It is thus possible to estimate the
final number of surviving particles by evaluating the solution to
Eq.~\eqref{rate-eq} when $V=N$ at time $t\sim T\sim N^{2/3}$.
According to the above definitions, $M_+= \displaystyle
\lim_{t\to\infty} n_+(t)$ and similarly, $M_-= \displaystyle
\lim_{t\to \infty}n_-(t)$. We anticipate that
\begin{equation}
\label{final}
M_+\sim n_+(T),\qquad M_-\sim n_-(T)\,.
\end{equation}
As discussed below, our numerical simulations confirm these behaviors
for a wide range of initial conditions.

\section{Equal Populations}

\begin{figure}[t]
\includegraphics[width=0.45\textwidth]{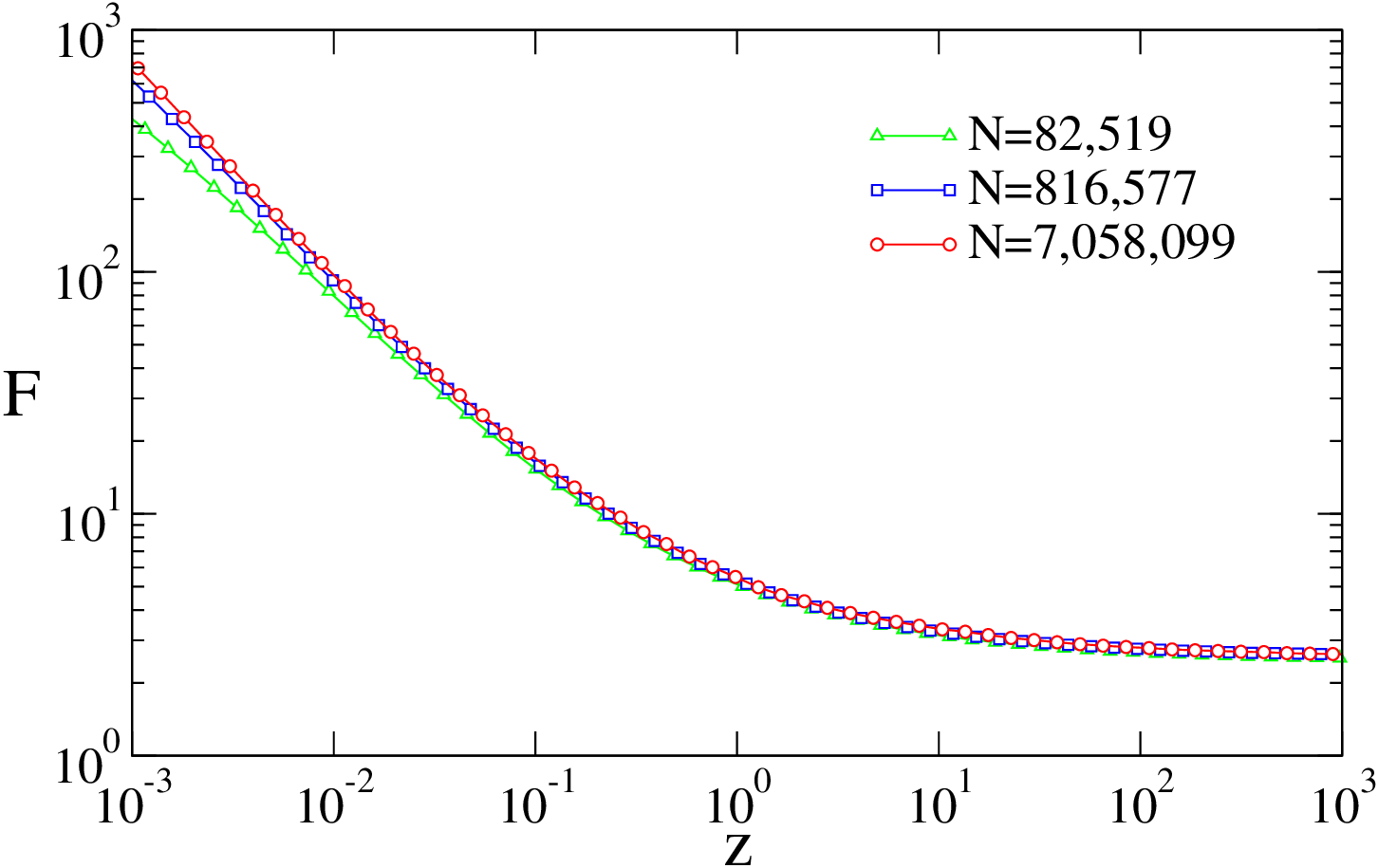}
\caption{The scaling function $F(z)$ defined in \eqref{scaling-n}.
  Shown is the scaled number of particles $F\equiv n/N^{1/3}$ versus
  the scaled time $z\equiv t/N^{2/3}$ for the case of equal number of
  particles for thee different system size.}
\label{fig-scaling-nt}
\end{figure}

We first consider the simplest case of equal populations, $\Delta=0$.
Since the number difference is conserved, the two populations are
identical $n_+=n_-=n/2$ with $n=n_++n_-$ the total population.  From
the rate equations \eqref{rate-eq}, the total population decays
according to
\begin{equation}
\label{n-eq}
\frac{dn}{dt} = - \frac{n^2}{2N},
\end{equation}
during the early phase $t\ll T$.  For the initial condition
$n(0)=N$, the population decays as the inverse of time,
\begin{equation}
\label{nt}
n(t)\sim N\,t^{-1}\,.
\end{equation}
Let $M= \displaystyle \lim_{t\to\infty}n(t)$ be the average number of
surviving particles.  At the crossover time, the number of particles
$M\sim n(T)$ is therefore (see figure \ref{fig-mnep})
\begin{equation}
\label{M}
M\sim N^{1/3}\,.
\end{equation}
Our numerical simulations, shown in figure \ref{fig-mnep}, confirm
this scaling relation. As expected, the vast majority of annihilation
events occur during the early phase when particles are inside the
initially occupied region. A finite fraction of the particles that
manage to survive at time $T$ persists forever.

The scaling relations \eqref{T} and \eqref{M} specify the typical
time scale and the typical surviving population. These scaling
laws fully characterize the time dependent behavior as the scaled
population $n/M$ is a universal function of the scaled time $t/T$ for
large systems (figure \ref{fig-scaling-nt})
\begin{equation}
\label{scaling-n}
n(t)\sim N^{1/3}F\left(t/N^{2/3}\right)\,,
\end{equation}
where $F(u) \sim u^{-1}$ for $u << 1$ and $F(u) \sim u^0$ for $u >> 1$. 

The average number of surviving particles \eqref{M} and the
finite-size scaling behavior \eqref{scaling-n} agree with the
corresponding behaviors for the single-species annihilation
process. Hence, a finite and equal number of particles from each
species survive the annihilation process when the initial populations
are identical.

\section{The Critical Difference}

In the rest of this study, we consider situations where the two
populations differ in size, $\Delta>0$. In this section, we analyze
the case where the population difference is fixed, that is, the
disparity between the two populations is always equal to $\Delta$.

\begin{figure}[t]
\includegraphics[width=0.45\textwidth]{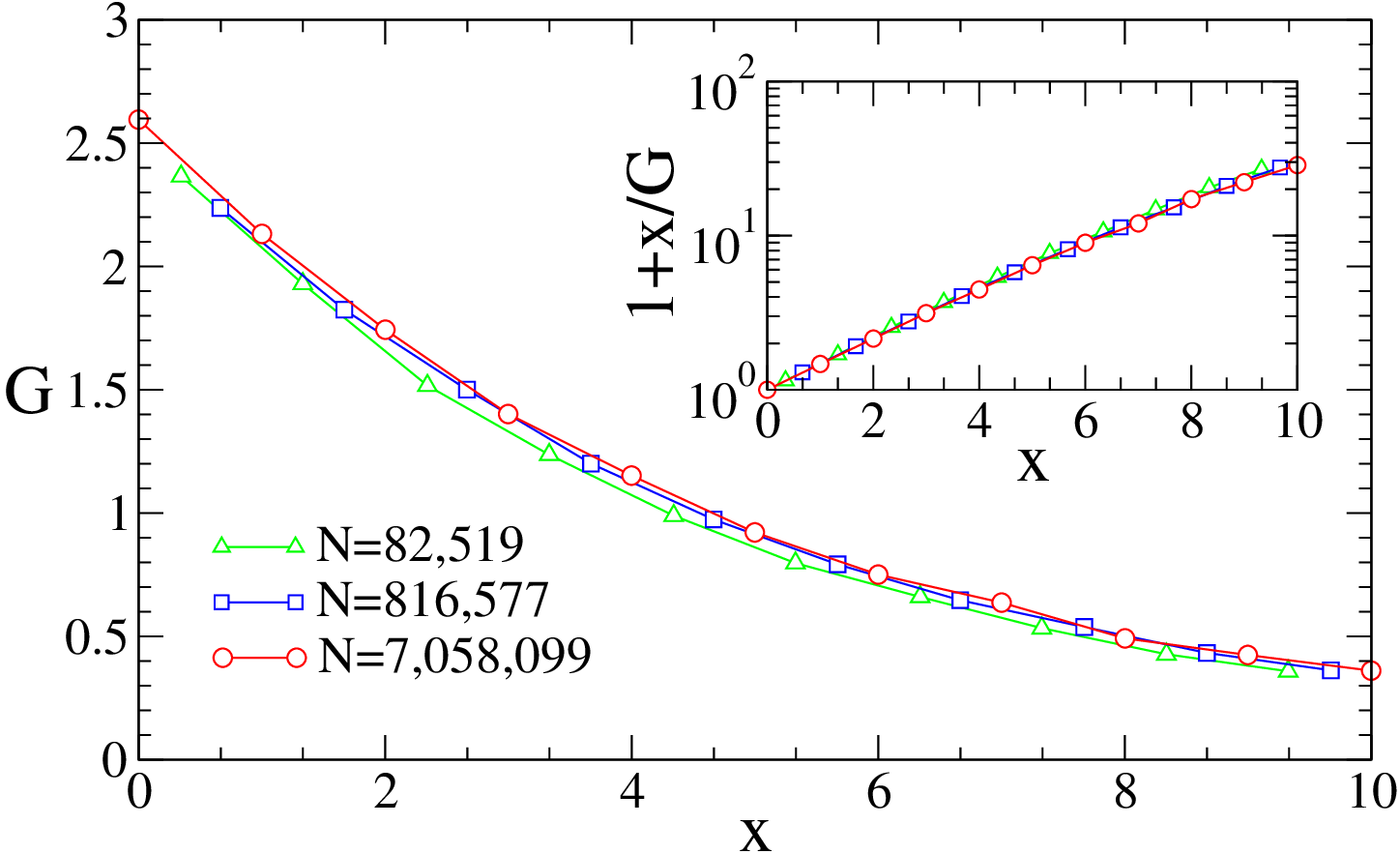}
\caption{The scaling function $G(x)$, defined in equation
  \eqref{Gx-def}. Shown is the scaled number of surviving minority
  particles $G\equiv M_-/N^{1/3}$ versus the scaled difference
  $x\equiv \Delta/N^{1/3}$ for the case of fixed population difference
  $\Delta$.  Inset shows the function $1+x/G$, which according to
  Eq.~\eqref{Gx} should increase exponentially with $x$.}
\label{fig-gx}
\end{figure}

Since the population difference is conserved, we may consider the
minority population without loss of generality. By substituting
$n_+=n_-+\Delta$ and $V=N$ into \eqref{rate-eq}, we see that the
minority population decays according to
\begin{equation}
\frac{dn_-}{dt} = - \frac{n_-(n_-+\Delta)}{N}\,.
\end{equation}
The solution of this equation subject to the initial conditions
$n_-(0)=N_-$ can be readily obtained, 
\begin{equation}
\label{nminus}
n_-(t)=N_-\,\frac{\Delta}{N_-(e^{t\Delta/N}-1)+\Delta}\,.
\end{equation}
Hereinafter, the dependence of $n_-(t)$ on $\Delta$ and $N$ is left
implicit.  We can recover the decay \eqref{nt} from \eqref{nminus} in
the limit $\Delta\to 0$.

The average number of surviving minority particles can be estimated by
evaluating the minority population \eqref{nminus} at the crossover
time \eqref{T},
\begin{equation}
\label{mminus}
M_-\sim N_-\,\frac{\Delta}{N_-(e^{\Delta/N^{1/3}}-1)+\Delta}\,.
\end{equation}
According to this expression, the number of surviving minority
particles grows algebraically with system size as in \eqref{M} when
$\Delta\ll N^{1/3}$, but it decays exponentially when $\Delta\gg
N^{1/3}$. Therefore, there is a critical difference, given by
\eqref{critical}, and drastically different behaviors occur above and
below this threshold,
\begin{equation}
\label{M-limits}
M_-\sim 
\begin{cases}
N^{1/3} & \Delta \ll \Delta_c, \cr
\Delta\,\exp\!\left(-c\,\Delta/N^{1/3}\right) & \Delta \gg
\Delta_c.
\end{cases}
\end{equation}
Here, $c$ is a constant.  For subcritical differences, a finite number
of minority particles survive and the same hold for the majority
species. Essentially, the system behaves as if the two populations are
equal. However, for supercritical differences, extinction of the
minority species is inevitable and the number of surviving majority
particles is always equal to the initial difference. In the $N\to\infty$
limit, we have $M_-\to 0$ and $M_+\to \Delta$.  Hence, the initial
disparity dictates if the minority species survives or if it becomes
extinct.  Also, the final number of surviving particles fluctuates in
the subcritical case but it is deterministic in the supercritical
case.

We note that in the supercritical region, $\Delta\gg \Delta_c$, there
is an additional characteristic time scale. Initially, the two
populations are comparable and consequently, the decay \eqref{nt}
holds. However, the two populations are no longer comparable,
$n_-(\tau)\sim \Delta$ at time $\tau\sim N/\Delta$.  The majority
population becomes dominant, $n_+\gg n_-$ when $t\gg \tau$, and
according to \eqref{rate-eq}, the minority population decays
exponentially, $dn_-/dt = -\Delta n_-/N$, thereby leading to the
exponential decay in \eqref{M-limits}.

\begin{figure}[t]
\includegraphics[width=0.45\textwidth]{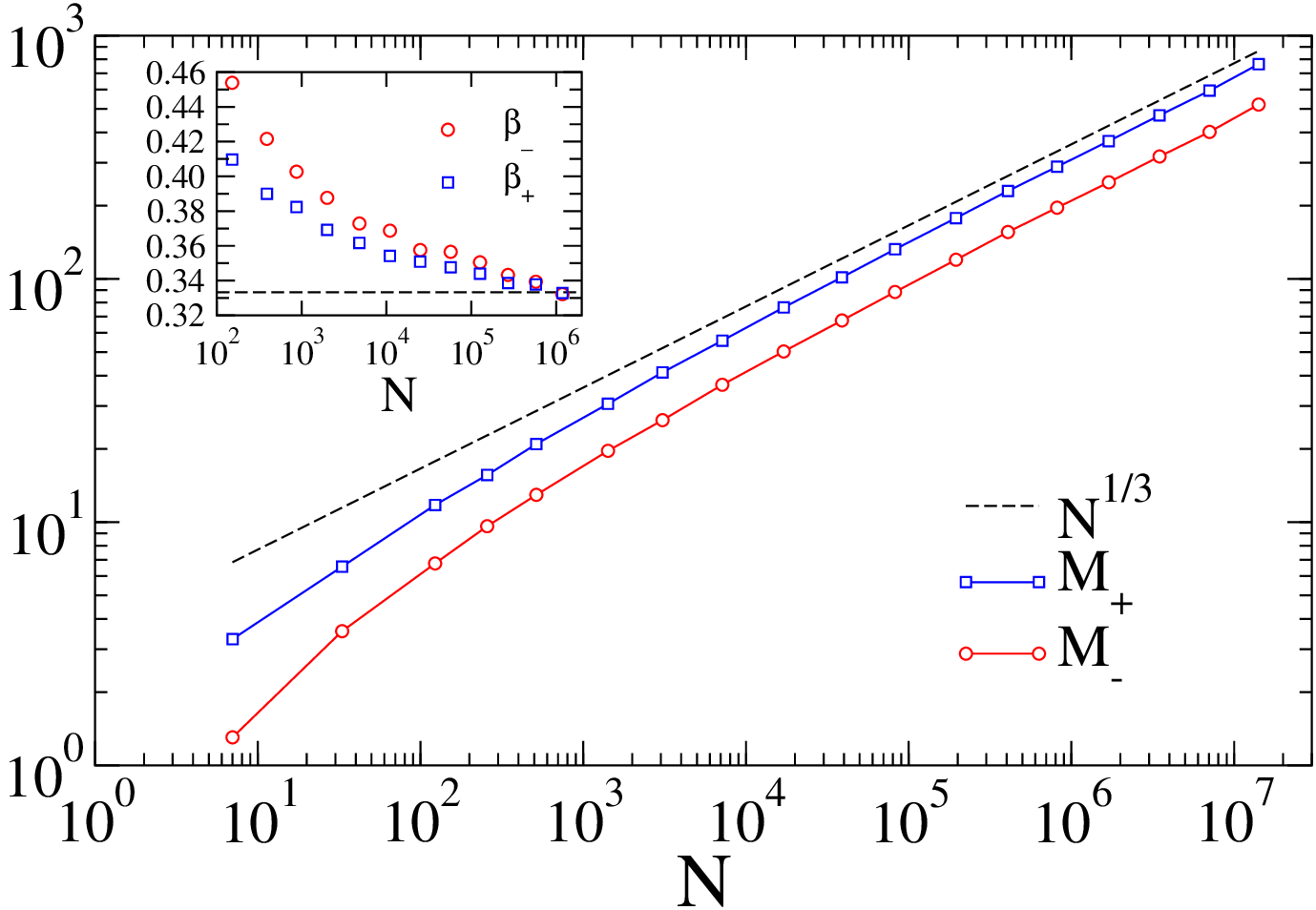}
\caption{The surviving populations $M_+$ and $M_-$ versus system size
  $N$ for the special case $\Delta=N^{1/3}$.  A fit of $M_+$ and $M_-$
  to a power-law, $M_\pm\sim N^{\beta_\pm}$ in the range $N>10^4$
  yields the exponents $\beta_+=0.34$ and $\beta_-=0.35$
  respectively. The inset shows the quantity $\beta_\pm\equiv 
d\ln M_\pm/d \ln N$ versus $N$.}
\label{fig-mncd}
\end{figure}

Numerically, we can confirm that the critical difference
\eqref{critical} characterizes the final population $M_-$. The
scaled number $M_-/N^{1/3}$ becomes a universal function of the scaled
difference $\Delta/N^{1/3}$ in the large-$N$ limit (figure
\ref{fig-gx})
\begin{equation}
\label{Gx-def}
M_-\sim N^{1/3}\,G\left(\Delta/N^{1/3}\right)\,.
\end{equation}
The underlying scaling function is simply
\begin{equation}
\label{Gx}
G(x)=\frac{x}{e^{c\,x}-1}\,.
\end{equation}
Results of our numerical simulations support this functional form as
well (see inset in figure \ref{fig-gx}).  The limiting behaviors
of the scaling function are
\begin{equation}
\label{Gx-limits} 
G(x)\sim 
\begin{cases}
1& x \ll 1 \cr
x\,e^{-c\, x} & x\gg 1\,.
\end{cases}
\end{equation}
The small-$x$ behavior shows that the problem reduces to the equal
population case in the subcritical regime. The large-$x$ behavior
reflects the extinction in the supercritical regime.

\begin{figure}[t]
\includegraphics[width=0.45\textwidth]{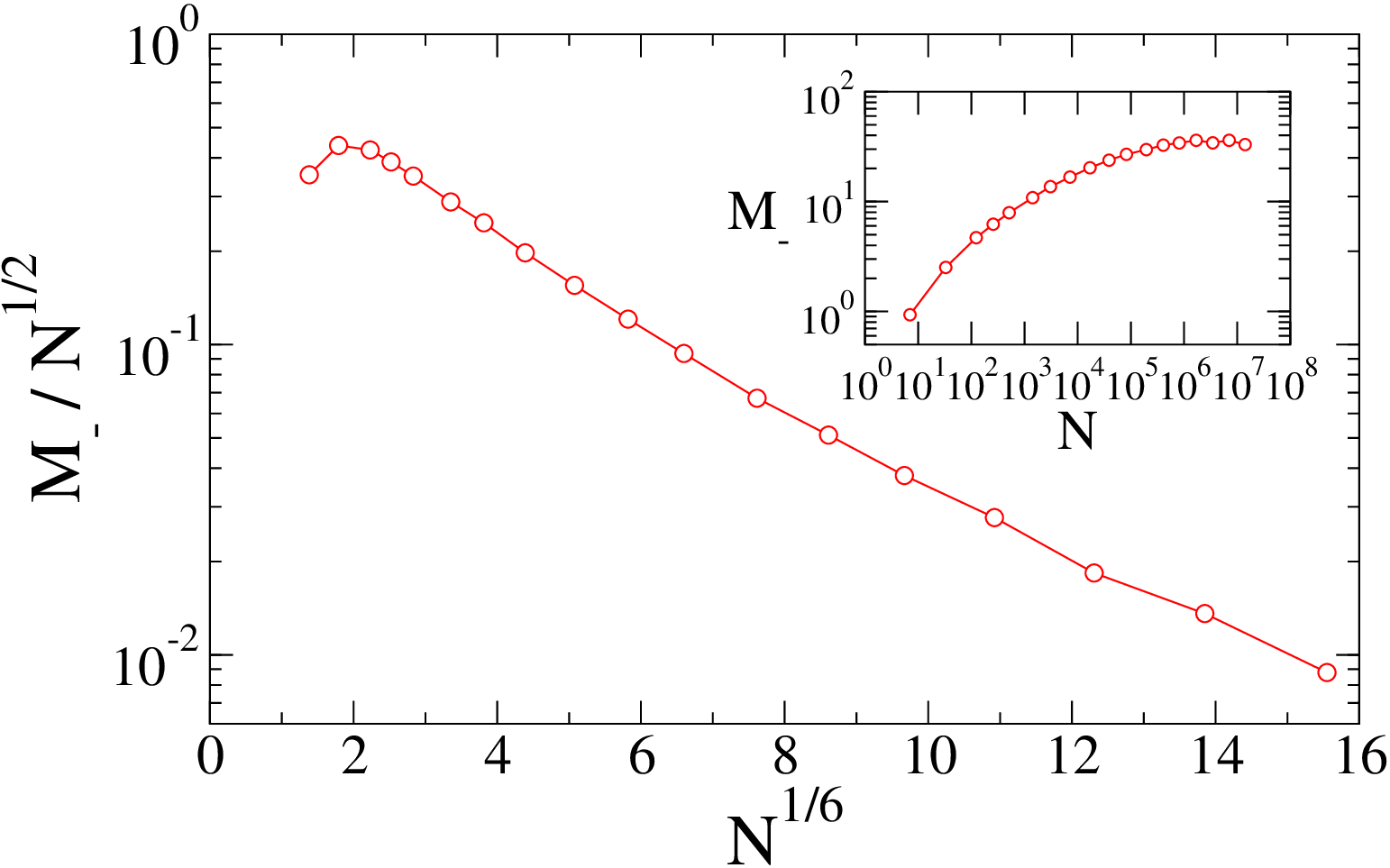}
\caption{The normalized population $M_-/N^{1/2}$ versus $N^{1/6}$ for
  the special case $\Delta=N^{1/2}$. The inset shows the surviving
  population $M_-$ versus system size $N$.}
\label{fig-mndet}
\end{figure}

To further verify the predictions of \eqref{mminus}, we examined two
special cases: $\Delta=N^{1/3}$ and $\Delta=N^{1/2}$. In the first
case, which corresponds to the critical difference, we can confirm that
$M_+\sim M_-\sim N^{1/3}$ (figure \ref{fig-mncd}). In the second case,
which is typical for equal initial concentrations, we expect a
stretched exponential decay with a rather small characteristic
exponent
\begin{equation}
\label{mminus1}
M_-\sim \sqrt{N}\,\exp\!\left(-c\,N^{\frac{1}{6}}\right)\,.
\end{equation}
Our numerical simulations are consistent with this behavior, despite
the fact that $M_-$ grows with system size over the range of system
sizes we probed numerically. 

\section{Equal Concentrations}

We now consider the situation where the initial concentrations are
equal. In this case, we have $N_+/N\to 1/2$ and $N_-/N\to 1/2$ in the
limit $N\to \infty$. The disparity between the two populations is a
fluctuating quantity, characterized by the typical difference $\Delta
\sim N^{1/2}$. Moreover, the difference is normally-distributed and
fully characterized by the distribution 
\begin{equation}
\label{pd}
P(\Delta) =\left(\frac{1}{2\pi N}\right)^{1/2}
\exp\left(-\frac{\Delta^2}{2N}\right)\,.
\end{equation}
We are interested in the average number of surviving particles $M_+$
and $M_-$, where the average is performed over all initial conditions
and all realizations of the annihilation process.

The scaling law for $M_+$ in \eqref{ec} follows from conservation of
the number difference. As discussed in Section II, the initial
difference provides a lower bound for the final number of majority
particles, $M_+\geq \Delta$. For equal concentrations, $\Delta\sim
N^{1/2}$, and according to \eqref{critical} the system is typically in
the supercritical regime. Consequently, the majority is dominant and
$M_+\sim N^{1/2}$.

The scaling law for $M_-$ can also be obtained using heuristic
arguments. According to equation \eqref{M-limits}, the minority
population disappears when $\Delta\gg N^{1/3}$, but some minority
particles do remain when $\Delta \ll N^{1/3}$. For equal
concentrations, initial conditions of the former type occur with high
probability, but initial conditions of the latter type may still be
realized with a small probability. To estimate this small probability
we conveniently replace the Gaussian in \eqref{pd} with a uniform
distribution with support in the compact interval $[-N^{1/2}:
  N^{1/2}]$. The initial difference is subcritical with probability
$\sim N^{-1/2}\times \Delta_c \sim N^{-1/6}$. Therefore, the average
size of the surviving minority population is $M_-\sim N^{-1/6}\times
N^{1/3}\sim N^{1/6}$.

Thus, there are two different scaling laws for majority and minority
survivors, $M_+\sim N^{1/2}$ and $M_-\sim N^{1/6}$. Yet neither of
these behaviors resembles the scaling behavior \eqref{ep} when the
populations are equal.  These two scaling relations give the survival
probability of a majority particle, $S_+\sim N^{-1/2}$, and that of
a minority particle, $S_-\sim N^{-5/6}$. The former survival
probability increases by a factor $\sim N^{1/6}$, while the latter
decreases by a similar factor when compared with the equal population
case where $S_+\sim S_-\sim N^{-2/3}$.

\begin{figure}[t]
\includegraphics[width=0.45\textwidth]{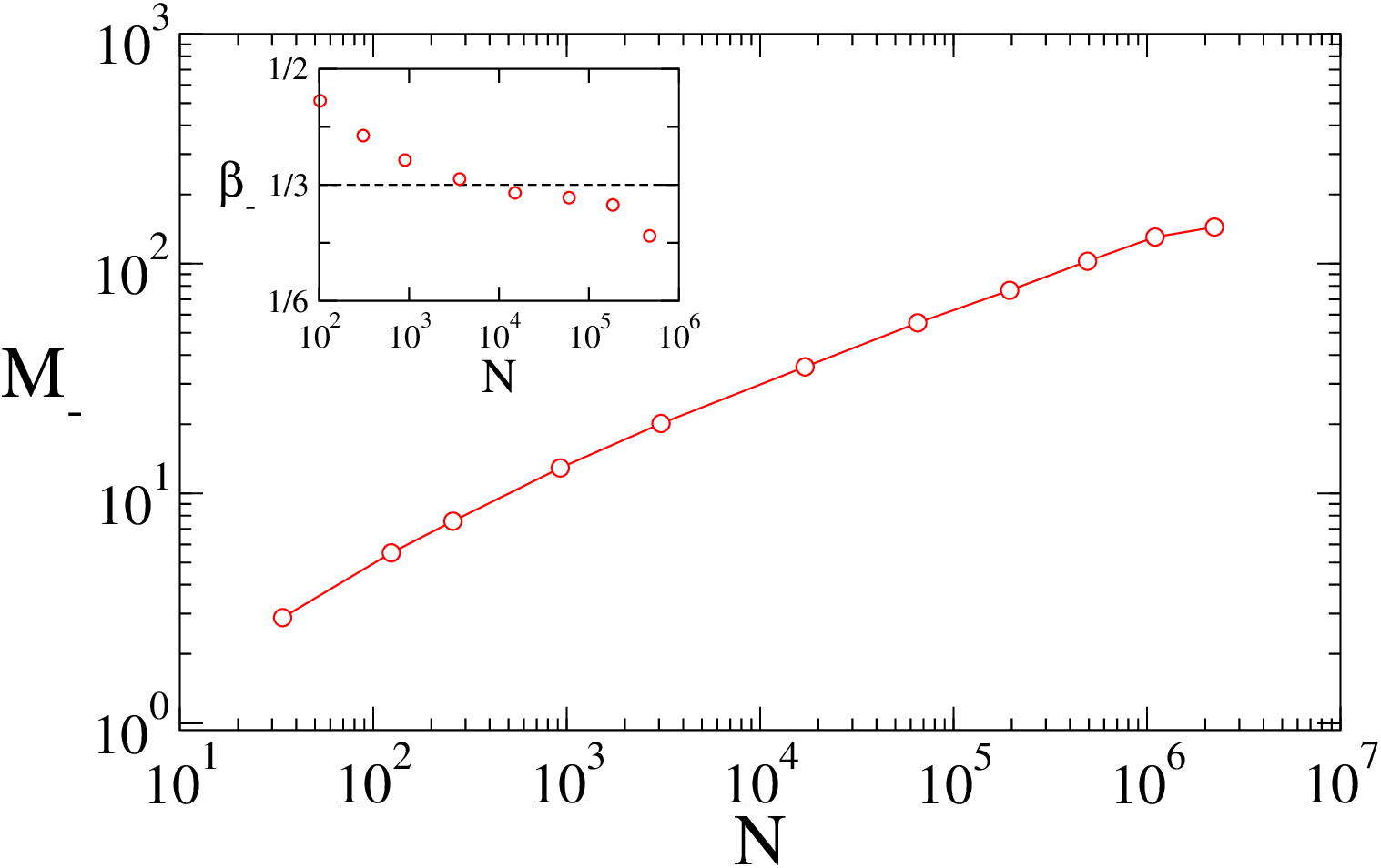}
\caption{The average number of surviving minority particles $M_-$ 
versus $N$ for the equal concentration case. The inset shows the quantity 
$\beta_-=d\ln M_-/d\ln N$ versus $N$.}
\label{fig-mnec}
\end{figure}

The surviving minority population may also be obtained by calculating
the weighted average $\int_0^\infty d\Delta P(\Delta) M_-$ with $M_-$
given in \eqref{mminus}. To estimate this integral, we use
\eqref{mminus} to separate contributions corresponding to the
subcritical phase and the supercritical phase,
\begin{eqnarray} 
\label{int}
M_-&\sim& 
\int_0^{N^{1/3}} d\Delta \left(\frac{1}{2\pi N}\right)^{1/2}\!\!N^{1/3}\,\exp\left(-\frac{\Delta^2}{2N}\right)\, \\
&+& 
\int_{N^{1/3}}^\infty d\Delta \left(\frac{1}{2\pi N}\right)^{1/2}
\!\!\Delta\,\exp\left(-\frac{\Delta^2}{2N}-\frac{\Delta}{N^{1/3}}\right)\,.\nonumber
\end{eqnarray} 
The first integral, which corresponds to the subcritical phase, is
much larger than the second one, and indeed it gives $M_-\sim
N^{1/6}$\,. Our numerical simulations show that $M_-\sim N^{\beta_-}$
with $\beta_-<1/3$ (figure \ref{fig-mnec}), but the exponent $\beta_-$
converges very slowly to the asymptotic value. While we are unable to
provide direct numerical evidence for $\beta_-=1/6$, when taken as a
whole, the rest of our simulation results do support this theoretical 
prediction.

We stress that the algebraic behavior $M_-\sim N^{1/6}$ characterizes
an average over all realizations of the stochastic process and over
all initial conditions. The initial difference $\Delta$ fluctuates
from realization to realization and it is governed by the distribution
\eqref{pd}.  Once the initial conditions are set, the fate of the
system is determined from the initial difference $\Delta$.  There is a
critical threshold $\Delta_c\sim N^{1/3}$. Below this threshold,
$\Delta\ll \Delta_c$, a finite number of particles from both the
majority and the minority survive ad infinitum and $M_-\sim M_+\sim
N^{1/3}$.  Above this threshold, however, all minority particles are
annihilated and a finite number of majority particles survive: $M_-\to
0$ and $M_+\sim \Delta$.

Clearly, there are wild fluctuations from realization to
realization. In the most probable scenario, the minority species goes
extinct, but there are rare cases where the minority species survives
and its population is comparable with that of the majority species.
One way to characterize these fluctuations is through moments of the
fluctuating number of minority survivors, $n_-(\infty)$.  It is simple
to generalize \eqref{int} and find a continuous spectrum of scaling
exponents that characterizes these moments 
\begin{equation}
\label{moments}
\langle n_-(\infty)^m\rangle \sim N^{\frac{2m-1}{6}}\,.
\end{equation}
The decaying zeroth moment reflects that initial conditions with
$\Delta\sim N^{1/3}$ are realized with probability $N^{-1/6}$.  The
behavior of large moments is controlled by the scaling law \eqref{ep}
for equal populations.

\section{Monte Carlo Algorithms} 

Numerical simulations of two-species annihilation with a finite, yet
large, number of particles are challenging for multiple
reasons. First, the system is three dimensional.  Even sophisticated
Monte Carlo algorithms \cite{sb}, developed recently, have not been
able to produce numerical verification of the decay $c\sim t^{-3/4}$
in unbounded systems because the convergence to the ultimate
asymptotic behavior is extremely slow \cite{bulatov}.  Second, large
memory is required because particles may escape far outside the
initially occupied domain.  Third, the computing time is large because
the very last annihilation event is unknown apriori and it fluctuates
greatly from one realization to another. Knowledge of the time scale
\eqref{T} is helpful with respect to this third challenge, and we run
our simulation to time $t_f$ much larger than this scale, $t_f \propto
10^4\times N^{2/3}$.

In our numerical simulations, $N$ sites that fall within a fixed
distance from the origin are occupied initially, but all remaining
lattice sites are vacant. In each elementary simulation step, one
randomly-selected particle moves to a randomly-selected nearest
neighbor. If the target site contains a particle of the opposite type,
the two particle are removed from the system. Time is augmented by the
inverse number of remaining particles after each such elementary
simulation step.  We used two different algorithms to simulate this
process. The two implementations differ in one respect only: in the
first algorithm we do allow multiple occupancy, but in the second, we
restrict occupancy to one particle per site. In the first
implementation each of the $N$ sites are occupied by one majority
particle and one minority particle but then $\Delta$ randomly-selected
minority particles are removed from the system. In the second
implementation each of the $N$ sites are occupied by a single
particle.  These two implementations yield very similar results which
become essentially indistinguishable for large systems.
 
\begin{figure}[t]
\includegraphics[width=0.45\textwidth]{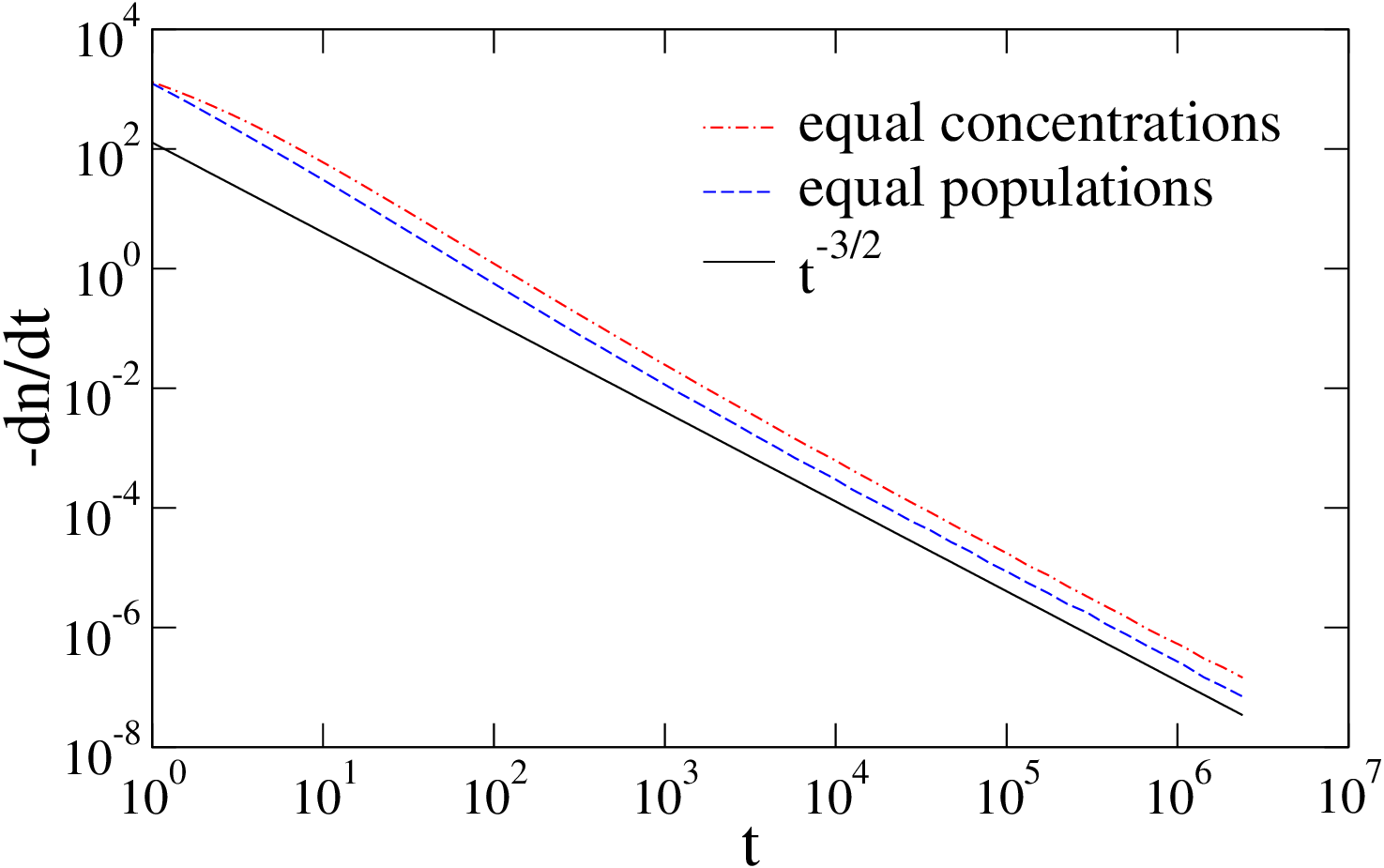}
\caption{The reaction rate $-dn/dt$ versus time $t$ for the case $N=7,153$.}
\label{fig-dndt}
\end{figure}

Our first simulation method is a brute force algorithm in which a
one-dimensional array is used to keep track of each particle's
location.  The advantage of this algorithm is that the memory required
scales linearly with the initial number of particles $N$.  However,
the number of operations per unit time grows quadratically with the
total number of active particles.  This algorithm performs
surprisingly well because most reaction events occur in the early
phase, and in particular, the number of required operations scales as
${\cal O}(N^2\ln N)$ in the subcritical phase \cite{bk}. We used this
straightforward algorithm to produce the results shown in figures
\ref{fig-mnep}-\ref{fig-mndet} and in figure \ref{fig-dndt}.

Our second algorithm is more sophisticated in that it is efficient in
both computation time and memory. To optimize the number of
operations, we implement the standard approach for simulating
diffusion-controlled reaction processes. Particles occupy an actual
three-dimensional lattice and each lattice site contains a ``pointer''
to the particle occupying it such that one does not need to search
through all particles in each move. With this approach, the number of
operations per unit time scales linearly with the number of active
particles. To optimize memory use, we take advantage of the fact the
system becomes sparse with time.  We thus map every lattice site in
our very large array to a much smaller array using a hash function
\cite{clrs, sa}.  This approach allows us to simulate a large system
with much less memory than would be needed if we stored the entire
original lattice, and yields a speed up of up to a factor $10$ for
$N\approx 10^6$. Results of this simulation algorithm are shown in
Fig.~\ref{fig-mnec}.

Our basic assumption, stated in equation \eqref{final}, is that the
number of surviving particles at time $T\sim N^{2/3}$ yields the
correct scaling behavior for the surviving populations. To further
test this assertion, we examined the reaction rate at late
times. According to the rate equation \eqref{rate-eq} and the
confining volume in \eqref{V}, we expect $dn_+/dt\sim
n_+n_-/t^{3/2}$. Our simulations confirm this behavior for both equal
populations and equal concentrations.  Hence, the residual correction
to the ultimate number of surviving particles decays algebraically,
$n(t)-M\sim t^{-1/2}$, and from this behavior it is simple to show that
the total number of reaction events in the late-time regime $t\gg
N^{2/3}$ is small enough so that \eqref{final} holds.

\section{General Spatial Dimensions}

We now briefly discuss the behavior in general spatial dimensions; we
restrict our attention to the equal concentration case and dimensions 
$d>2$ where the final state is nontrivial.  It is straightforward to
generalize the main results \eqref{critical} and \eqref{ec} to
arbitrary dimension by replacing the characteristic time scale in
\eqref{T} with $T\sim N^{2/d}$.  The critical difference grows
algebraically with the number of particles,
\begin{equation}
\Delta_c\sim N^\delta \qquad{\rm with}\qquad \delta=
\frac{d-2}{d} \,,
\end{equation}
when $d>2$.  The surviving majority population exhibits two regimes of
behavior
\begin{equation}
M_+\sim N^{\beta_+},\qquad
\beta_+=
\begin{cases}
\frac{1}{2} & d\leq 4\cr
\frac{d-2}{d} & 4\leq d\,.\cr
\end{cases}
\end{equation}
The behavior agrees with \eqref{ec} below the critical dimension, and
the behavior coincides with that of single-species annihilation above
the critical dimension.  Finally, the surviving minority population
exhibits three regimes of behavior
\begin{equation}
\label{beta}
M_-\sim N^{\beta_-}, \qquad \beta_-=
\begin{cases}
0 & d\leq \frac{8}{3},\cr
\frac{3d-8}{2d} & \frac{8}{3}\leq d\leq 4, \cr
\frac{d-2}{d} & 4\leq d\,.\cr
\end{cases}
\end{equation}
Interestingly, the quantity $M_-$ does not grow with system size below
the lower critical dimension $d<8/3$ \cite{kr}.  However, the two
surviving populations are comparable, and both are much larger than
the initial difference $\Delta \sim N^{1/2}$ above the upper critical
dimension, $d>4$.

\section{conclusions}

To summarize, we studied diffusion-controlled two-species annihilation
in an unbounded space with a finite number of particles. Specifically,
we addressed initial conditions where a finite number of particles is
confined to a compact domain. We found that the disparity between the
two populations controls the behavior.  When the disparity is small
enough, the two populations remain comparable throughout the reaction
process, and a finite number survives the annihilation process. These
particles manage to escape far outside the initial domain. However,
when the initial disparity is large enough, the minority population
becomes extinct while a finite number of majority particles
survives. We used the rate equation approach to obtain a number of
scaling laws for equal initial populations and for equal initial
concentrations. Our numerical simulations support the theoretical
predictions.

Our study focused on the most interesting case of three dimensions
which is below the critical dimension $d_c=4$ for a homogeneous
infinite-particle systems \cite{krb}. For such systems, spatial correlations
spontaneously develop and the result is a mosaic of $A$-rich and
$B$-rich domains. These correlations play a crucial role and
consequently, predictions based on the mean-field rate equations do
not hold in three dimensions. The qualitative behavior is quite
different when the number of particles is finite. No matter how large
the initial number of particles is, the system remains well-mixed and
spatial correlations are not strong enough to affect the scaling
behavior. As a result, the rate equation predictions hold for finite
systems \cite{fds}.

Survival occurs only when $d>2$ and in some sense this escape
phenomena is counter to the behavior when the number of particles is
infinite. In an infinite system, the reaction rate is smaller in {\em
  low dimensions} whereas in a finite system, the total number of
reaction events is smaller in {\em high dimensions}. Hence, the system
size and dimensionality generally compete in reaction-diffusion
processes as both affect the survival probability.

Our study highlights the serious challenge of developing theoretical
tools for describing strongly interacting particle systems such as
reaction-diffusion processes involving a large but finite number of
particles.  Existing theoretical methods are inadequate to handle such
problems. As a rather straightforward extension of our work one may
investigate two-species annihilation with unequal initial
concentrations where according to \eqref{M-limits}, the survival
probability of minority particles decays as a stretched exponential, 
$S_-\sim \exp(-{\rm const.}\times N^{2/d})$. Finally, we mention that
it would be interesting to investigate another basic
reaction-diffusion process, Brownian coagulation \cite{mvs,sc}, for
initial conditions with a finite number of clusters.

\end{document}